\documentclass[fp,twocolumn]{jpsj3}
\usepackage{amsmath,amssymb}
\def\v#1{\mib #1}

\newcommand{\bra}[1]{\left\langle {#1} \right\vert}
\newcommand{\ket}[1]{\left\vert {#1} \right\rangle}
\def\sed{S_{\rm e}}
\def\JA{J_{\rm A}}
\def\JF{J_{\rm F}}
\def\Jl{J_{\rm L}}

\title
{
Characterization of Topological Phases of Spin-1/2 Frustrated Ferromagnetic--Antiferromagnetic Alternating Heisenberg Chains 
 by Entanglement Spectrum
}

\author
{
Kazuo { Hida}\thanks{E-mail address: hida@mail.saitama-u.ac.jp}}

\inst
{Division of Material Science, Graduate School of Science and Engineering, \\ Saitama University, Saitama, Saitama, 338-8570\\ 
}

\recdate
{
\today
}

\abst
{
The topological classification of a series of frustration-induced spin-gap phases in the  spin-1/2 ferromagnetic--antiferromagnetic alternating Heisenberg chain with  next-nearest-neighbour interaction reported in  J. Phys. Soc. Jpn. 82, 064703 (2013) is confirmed using  two kinds of entanglement spectra defined by different divisions of the whole chain. For the numerical calculation, the iDMRG method is used. The results are consistent with the valence bond solid picture proposed in the previous paper.}

\begin{document}
\sloppy
\maketitle 
\section{Introduction}
Since the discovery of the Haldane phase in spin-1 Heisenberg antiferromagnetic chains,\cite{hal1,hal2}  
the spin-gap phases in the ground states of quantum spin chains have been 
extensively studied in theoretical condensed matter physics. Recently, the spin-1 Haldane phase has been 
attracting renewed interest as one of the symmetry-protected topological (SPT)  phases of matter\cite{Pollmann2010,Pollmann2012,Zang2010,Hirano2008,Chen2011}. It is found that this phase is protected by space inversion, time reversal, and $\pi$-rotations around two axes.\cite{Pollmann2010,Pollmann2012}  From this viewpoint, the spin-gap phases of the quantum spin chains are classified into SPT and trivial phases in the presence of appropriate symmetry. 
It has also been proposed that these two phases can be distinguished by the even-odd parity of the degeneracy of their entanglement spectra (ES).\cite{Pollmann2010,Pollmann2012}

In the spin-1/2 frustrated ferromagnetic--antiferromagnetic alternating chain with next-nearest-neighbour interaction, the present author and coworkers found successive phase transitions in which a series of trivial and topological spin-gap phases alternate with the strength of frustration\cite{hts}. Such successive transitions between  a series of different spin-gap phases are rather unusual and nontrivial. 
 The exact solution is available on the ferromagnetic--nonmagnetic phase boundary.\cite{dmitriev,dmitriev2} In the previous paper\cite{hts}, we identified each phase from the number of additional $S=1/2$ spins necessary to compensate the edge spins. This method is, however, based on a rather intuitive physical argument. Also, there remained ambiguity in the strength of coupling between the additional spins and the chain. 
 It is the purpose of this work to confirm that these phases are actually characterized by the degeneracy of the ES that is a well-founded bulk property\cite{Pollmann2010,Pollmann2012}.

This paper is organized as follows. In the next section, the model Hamiltonian is presented and the ground-state phase diagram obtained in Ref. \citen{hts} is reviewed. In sect. 3, the entanglement spectrum is introduced and numerical results are presented.  The last  section is devoted to summary and discussion.

\section{Model and Ground State Phase Diagram}
\begin{figure}[h!]
\centerline{\includegraphics[width=7cm]{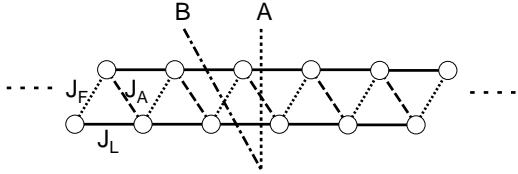}}
\caption{Lattice structure of the present system. In the calculation of the ES, the whole system is divided into left and right subsystems, as indicated by the dotted line (division A) and dash-dotted line (division B).}\label{fig:cut}
\end{figure}

We consider the $S=1/2$ Heisenberg chains described by the Hamiltonian
\begin{align}
{\cal H} &=\sum_{l=1}^{L} \left(\JF\v{S}_{2l-1}\v{S}_{2l}+\JA\v{S}_{2l}\v{S}_{2l+1}\right)\nonumber\\
&+\sum_{i=1}^{2L}\Jl\v{S}_{i}\v{S}_{i+2}, \label{hama}
\end{align}
where $\v{S}_{i}$  is the spin-1/2 operator on the $i$-th site. In this work, we focus on the case $\JF , \Jl < 0$  and $\JA > 0$. The lattice structure is shown in Fig. \ref{fig:cut}. This model has the same symmetries as the spin-1 Heisenberg chain, if  pairs of neighbouring spins are regarded as building blocks of the chain. Hence, the same classification of spin-gap phases should apply.

\begin{figure}
\centerline{\includegraphics[width=7cm]{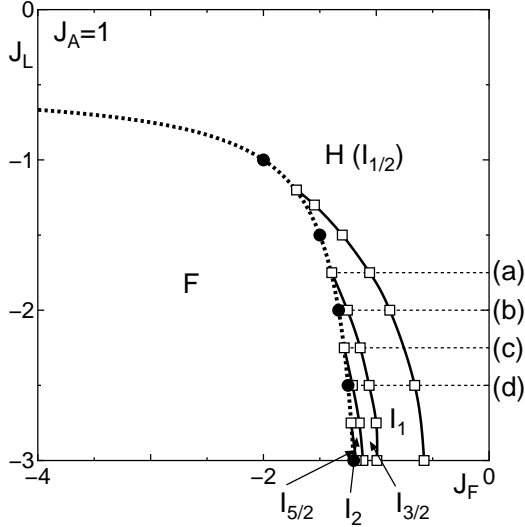}}
\caption{Ground state phase diagram. Phase boundaries are taken from Ref. \citen{hts}.  The ferromagnetic phase, the Haldane phase, and the intermediate spin-gap phases with edge spin $\sed$ in the open chain are indicated by F, H, and I$_{\sed}$, respectively. The dotted line is the stability limit of the ferromagnetic phase. The open squares are the boundary between different I$_{\sed}$ phases determined in Ref. \citen{hts}. The filled circles are the ``special points'' defined by Dmitriev {\it et al.}\cite{dmitriev,dmitriev2} with exact solutions. The solid curves are  guides for the eye.
 The entanglement spectra are calculated along the horizontal dotted lines (a)--(d).}
\label{fig:phase}
\end{figure}

Figure \ref{fig:phase} shows the ground-state phase diagram obtained in Ref. \citen{hts}. Between the Haldane (H) phase and the ferromagnetic (F) phase, a series of spin gap phases (I$_{\sed}$) are present. In the I$_{\sed}$ phase, the edge spins with magnitude $\sed$ appear at the two ends of the open finite chain expressed by the Hamiltonian (\ref{hama}). Each of these edge spins is compensated by the addition of $2\sed$  spins with magnitude $1/2$ on the edge.  

\section{Entanglement Spectrum}

To calculate the entanglement spectrum, we divide the whole chain into left and right subsystems. 
The density matrices of the right and left subsystems $\rho_{\rm R}$ and $\rho_{\rm L}$ are defined by
\begin{align}
\rho_{\rm R} &= \Tr_{\rm L}\ket{G}\bra{G}\\
\rho_{\rm L} &= \Tr_{R}\ket{G}\bra{G},
\end{align}
where $\ket{G}$ is the ground state of the whole system and $\Tr_{\rm L(R)}$ implies the trace over the left (right) subsystem. The  eigenvalues $w_{\alpha}$  and eigenstates $\ket{\alpha}_{\rm R(L)}$ of the density matrix $\rho_{\rm R (L)}$ satisfy the eigenvalue equations
\begin{align}
\rho_{\rm R}\ket{\alpha}_{\rm R} &= w_{\alpha}\ket{\alpha}_{\rm R} \\
\rho_{\rm L}\ket{\alpha}_{\rm L} &= w_{\alpha}\ket{\alpha}_{\rm L}.
\end{align}
The set of eigenvalues $\{w_{\alpha}\}$ is common to $\rho_{\rm R}$ and $\rho_{\rm L}$. They form an ES of the ground state of the whole system. Note that the ES depends on how the whole system is divided into subsystems as described below.

We employ the iDMRG\cite{McCulloch2008,Schllwock2011} method to calculate the ES. In this method, the matrix product wave function converges to that of the infinite size fixed point. 
Hence, the effect of the open edge is not reflected in the ES. Instead, the degeneracy of the ES 
results from the virtual free  spins that appear at the boundary between the left and right subsystems. Hence, this degeneracy of the ES is a bulk property.

\begin{figure}[h!] 
\centerline{\includegraphics[width=6cm]{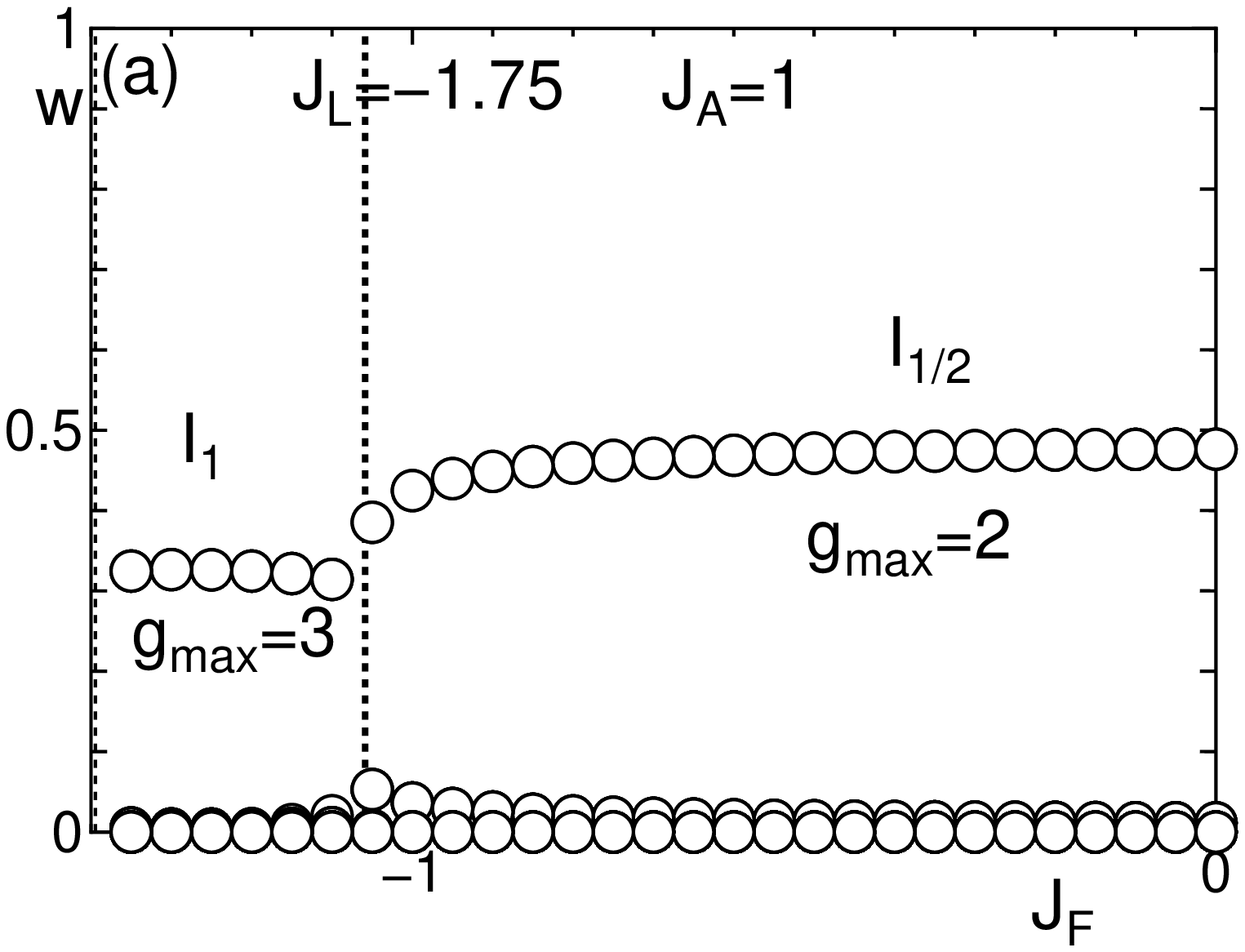}}
\centerline{\includegraphics[width=6cm]{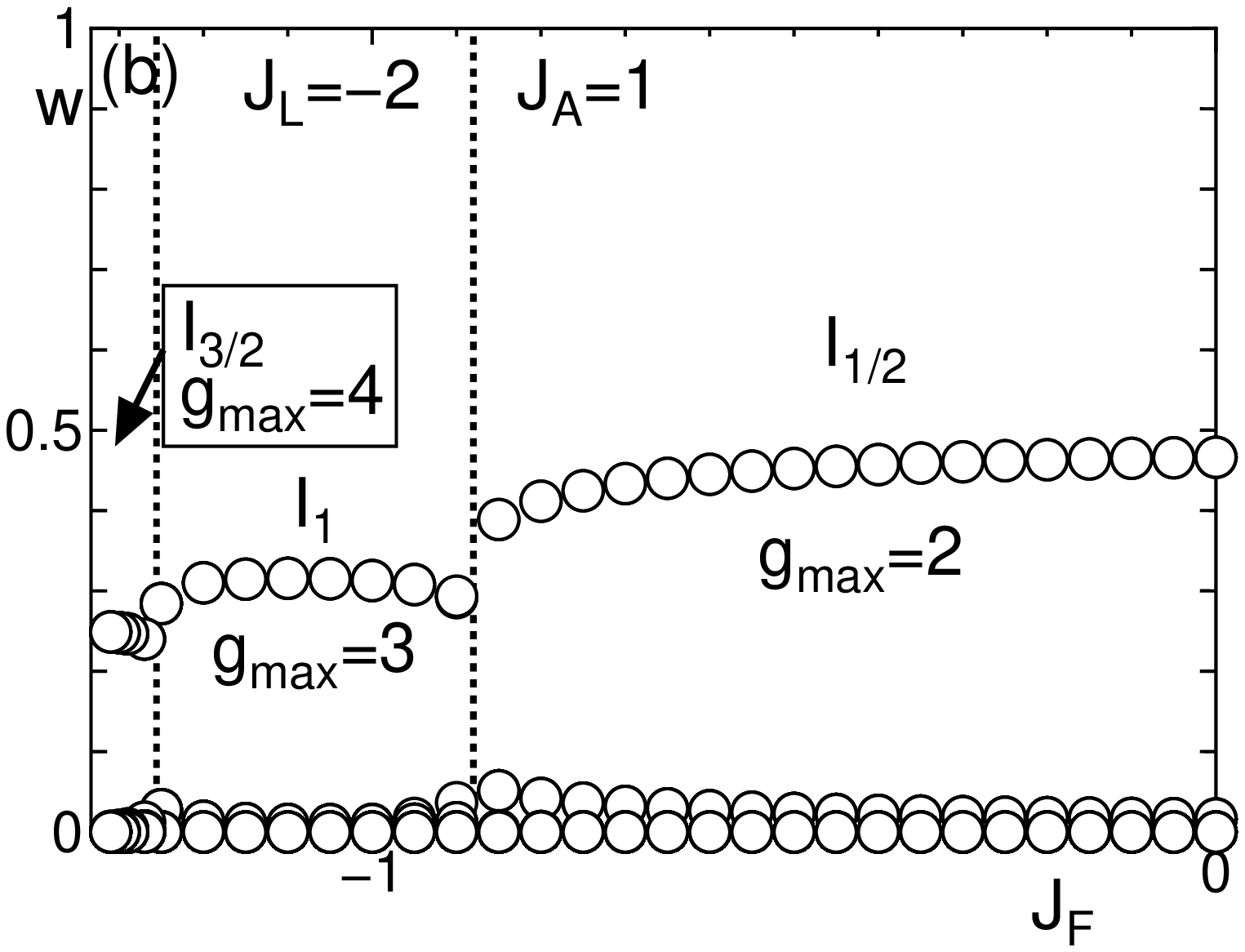}}
\centerline{\includegraphics[width=6cm]{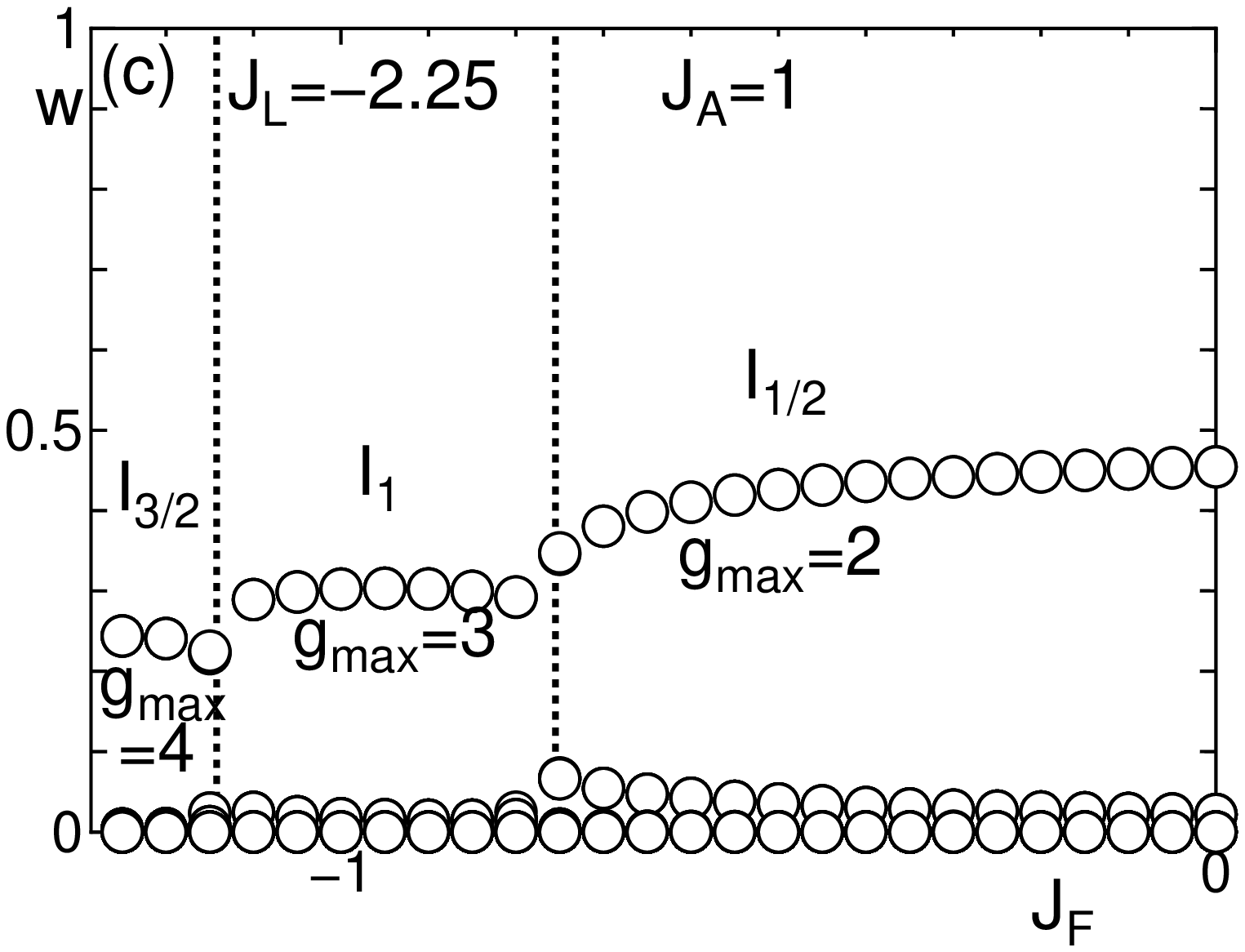}}
\centerline{\includegraphics[width=6cm]{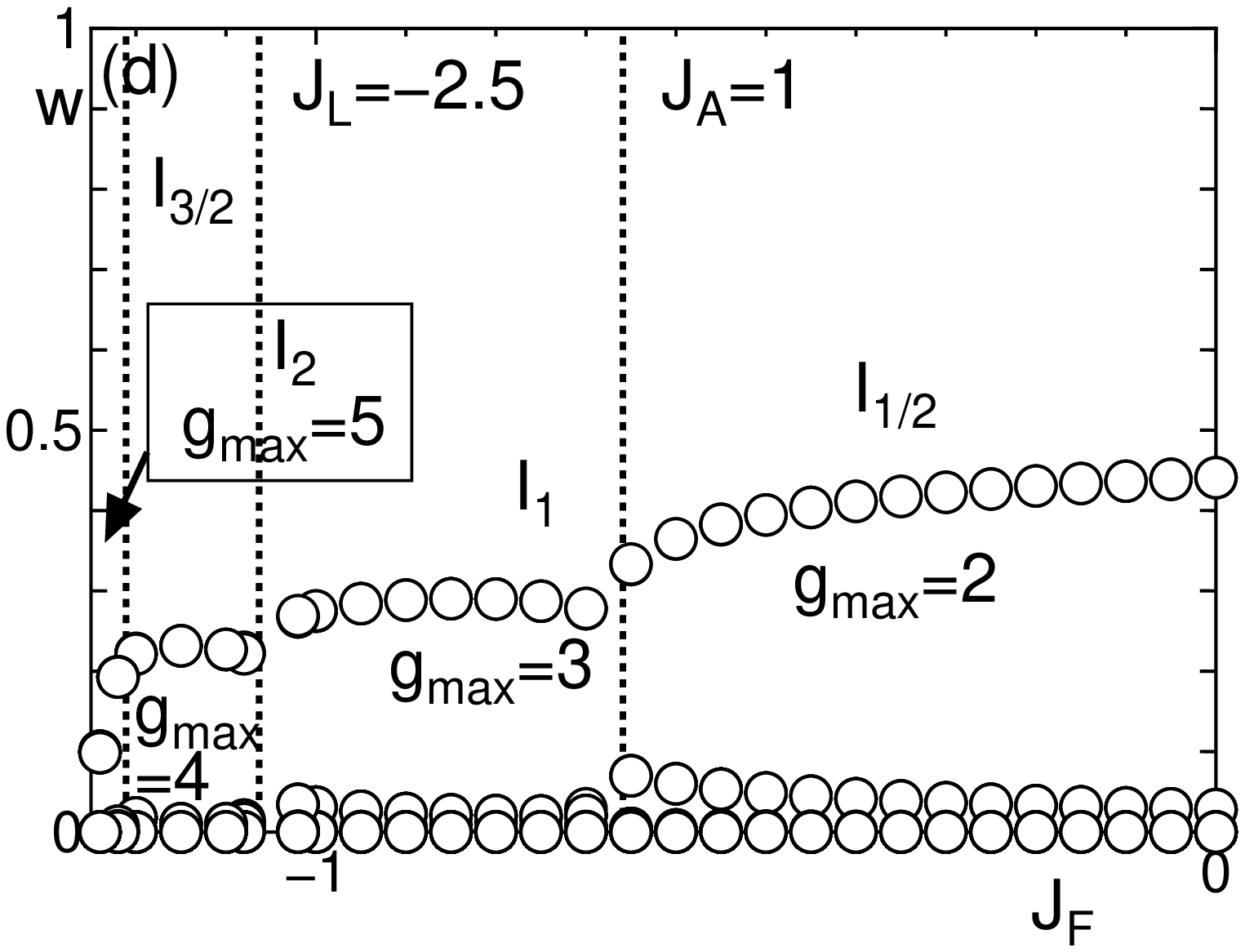}}
\caption{$\JF$-dependences of several large eigenvalues $w_{\alpha}$ of the density matrices $\rho_{\rm R(L)}$ and the degeneracy of the largest eigenvalue $g_{\rm max}$ with division A 
for (a) $\Jl= -1.75$, (b) $-2$, (c) $-2.25$, and (d) $-2.5$ with $\JA=1$. The vertical dotted lines are the phase boundaries determined in Ref. \citen{hts}. 
} 
\label{fig:enta}
\end{figure}

\begin{figure}[h] 
\centerline{\includegraphics[width=6cm]{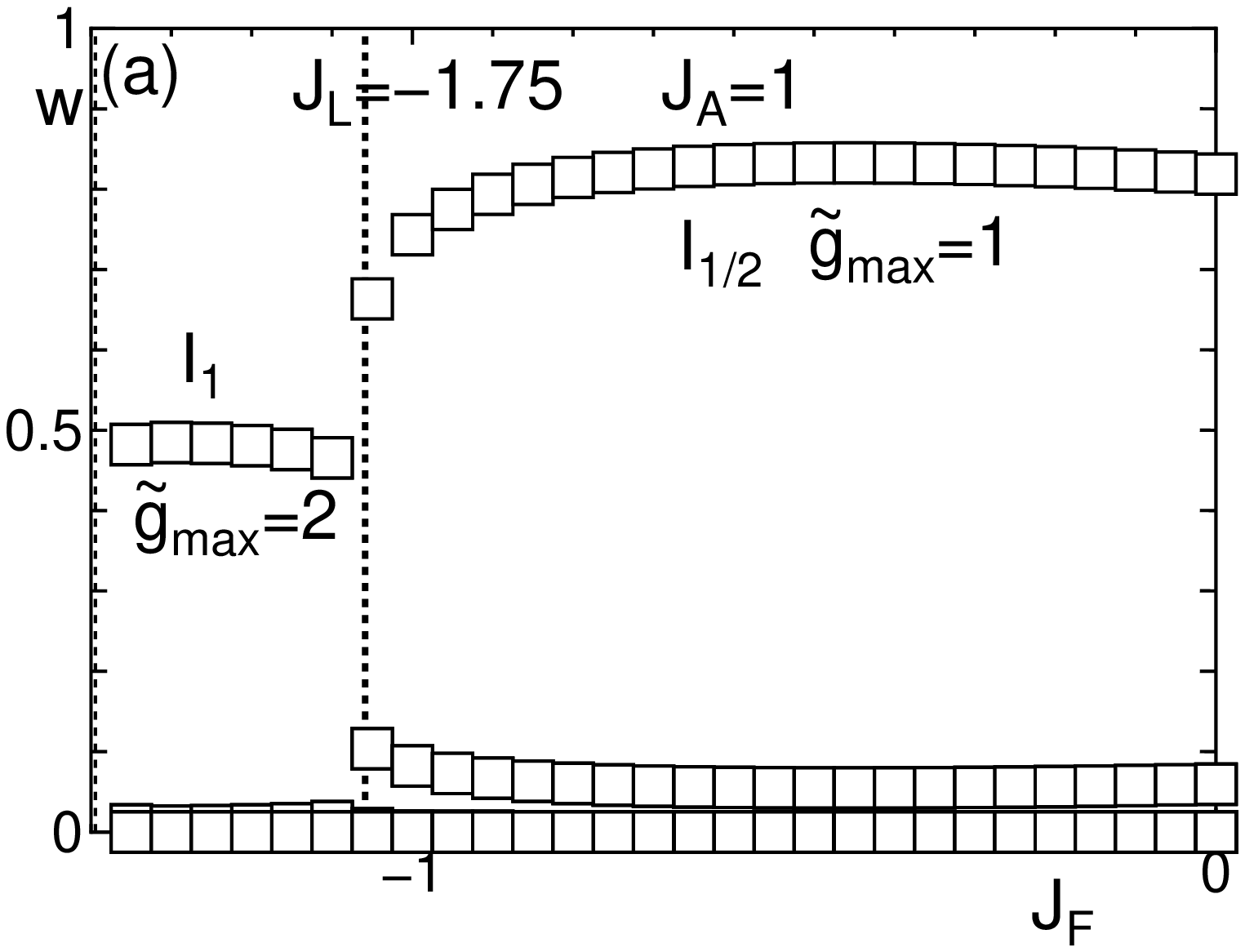}}
\centerline{\includegraphics[width=6cm]{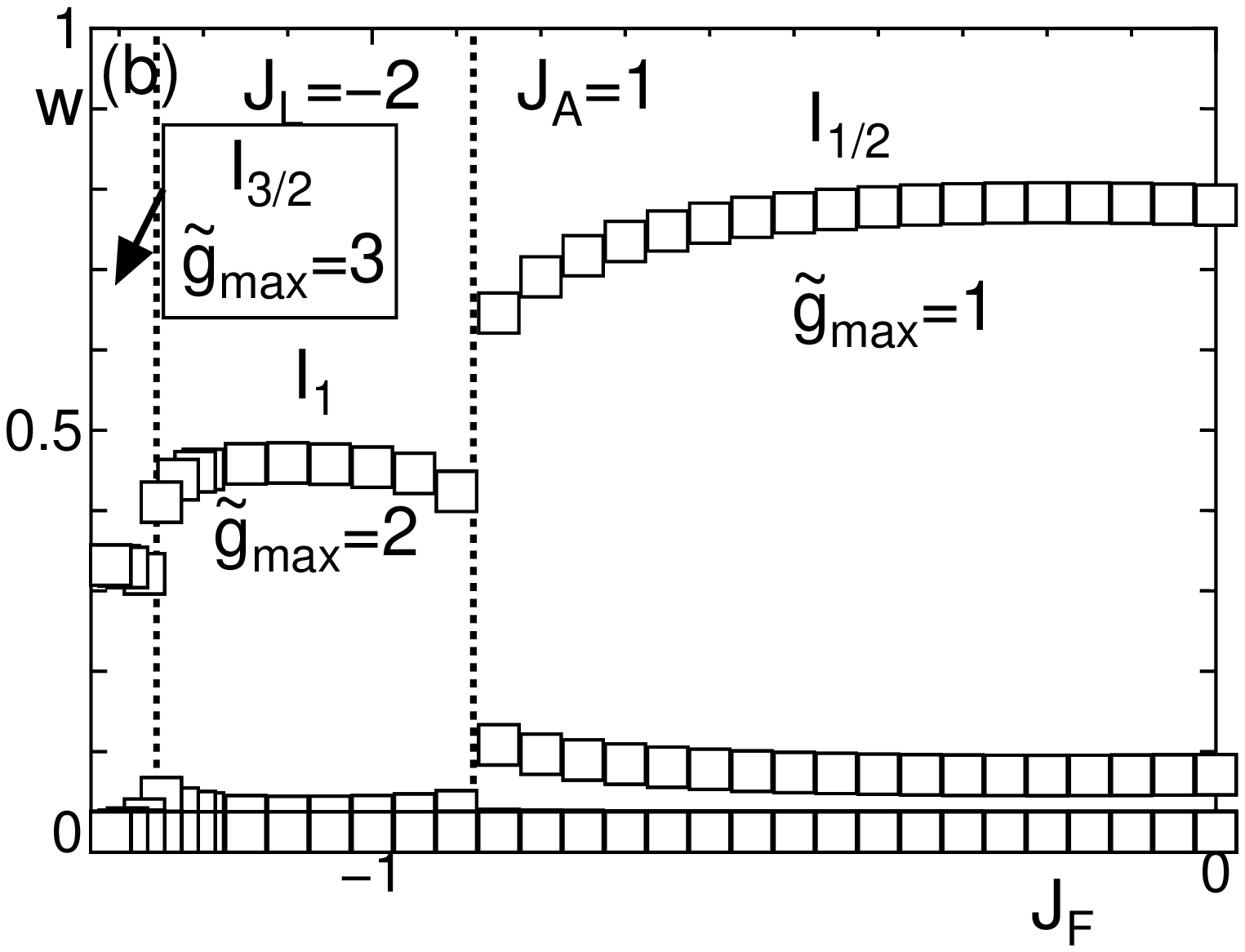}}
\centerline{\includegraphics[width=6cm]{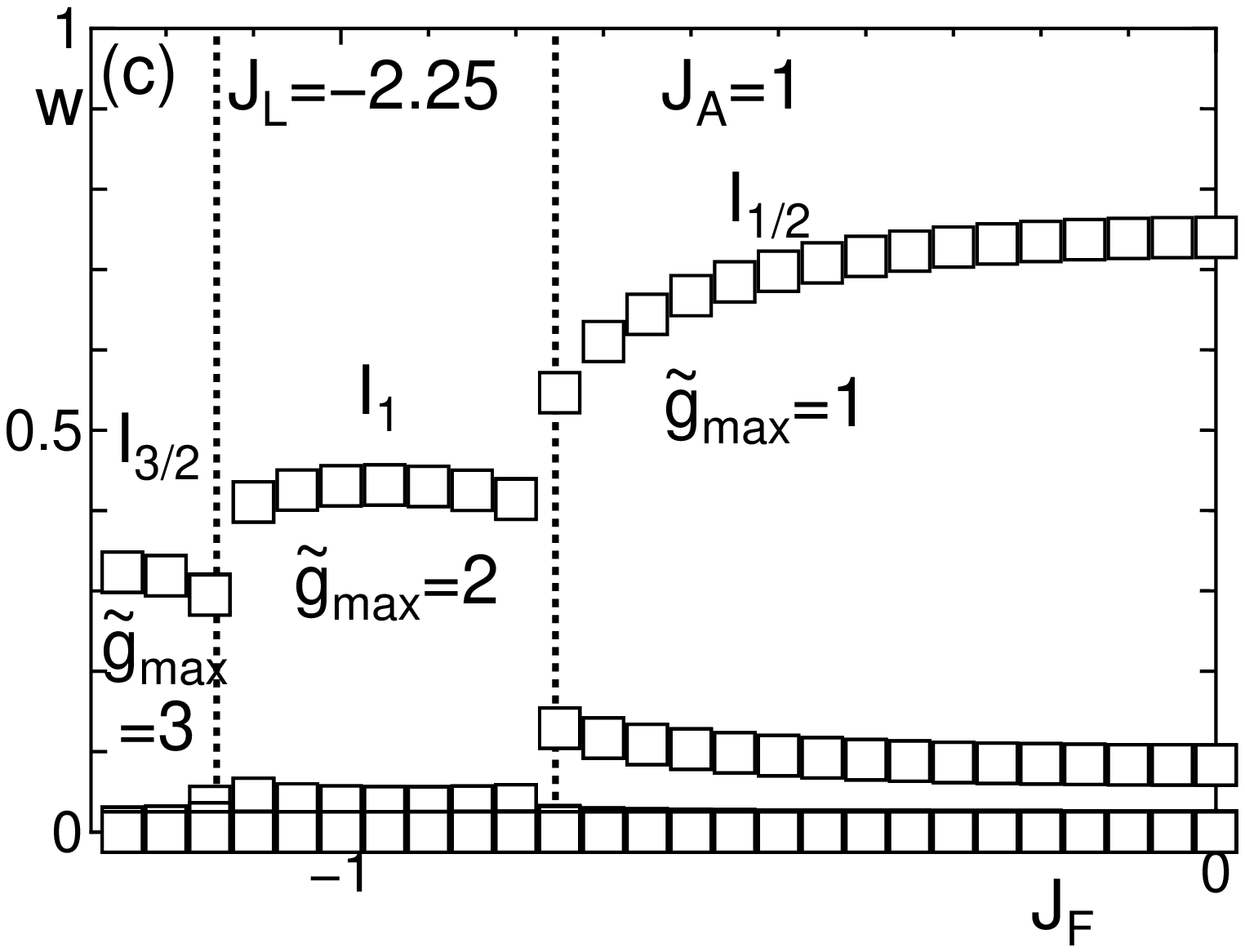}}
\centerline{\includegraphics[width=6cm]{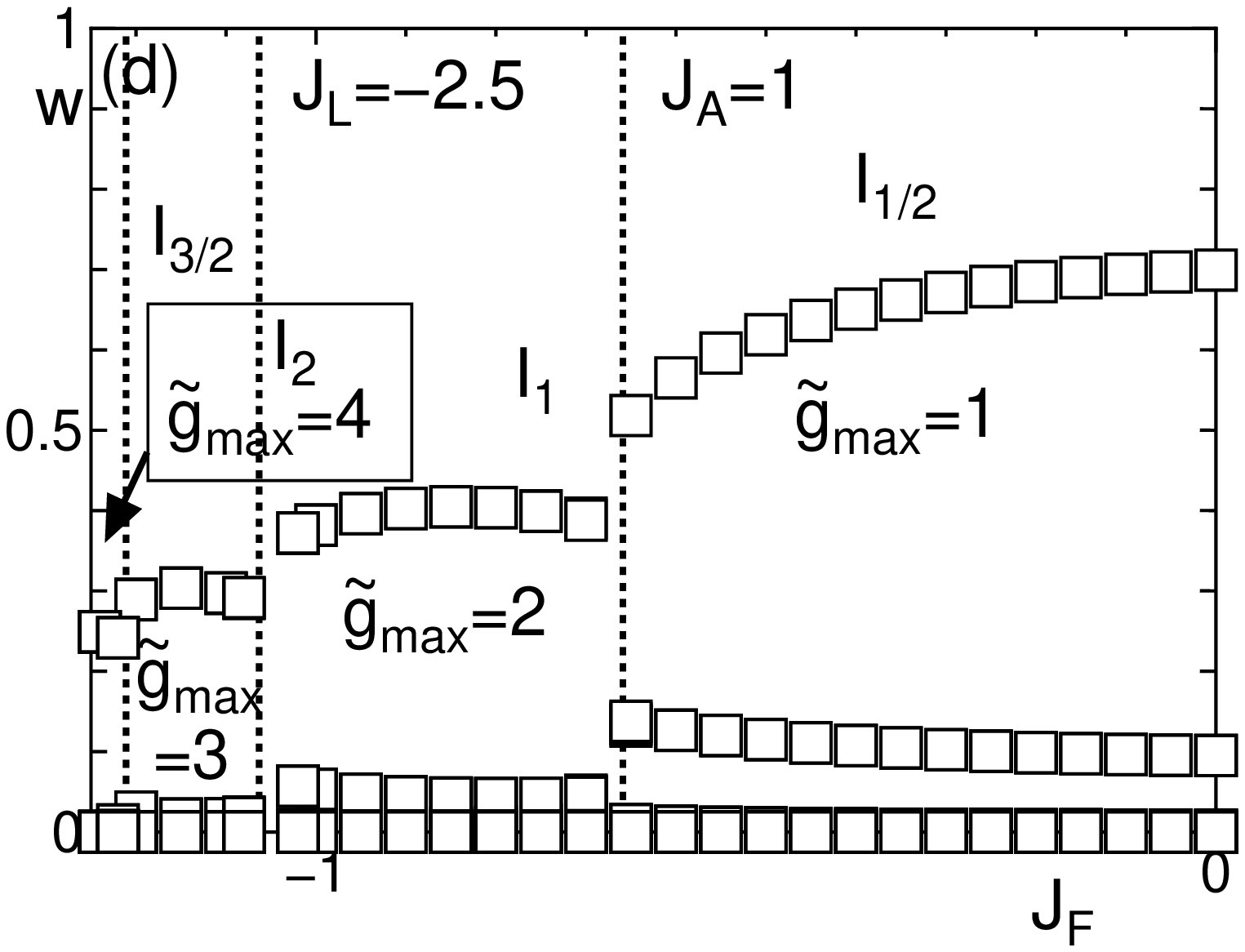}}
\caption{$\JF$-dependences of several large eigenvalues $w_{\alpha}$ of the density matrices $\rho_{\rm R(L)}$ and the degeneracy of the largest eigenvalue $\tilde{g}_{\rm max}$  with division B 
for (a) $\Jl= -1.75$, (b) $-2$, (c) $-2.25$, and (d) $-2.5$ with $\JA=1$. The vertical dotted lines are the phase boundaries determined in Ref. \citen{hts}. 
} 
\label{fig:entb}
\end{figure}

As we pointed out in Ref. \citen{hts}, the definitions of the terms ``topological'' and ``trivial'' are rather arbitrary in the present model.   
In the presence of one of the three symmetries that protect the Haldane phase, the ground states of the spin chains are classified into those with even and odd degeneracies of the ES. The spin-1 Haldane phase, which is unambiguously an SPT state, belongs to the former.\cite{Pollmann2010} Our model tends to the spin-1 chain in the limit $\JF \rightarrow -\infty$, where $\v{S}_{2l-1}$ and $\v{S}_{2l}$ form a spin-1 operator. In the absence of frustration, the ground state of the present model is the Haldane phase if the spin pairs connected by $\JF$ bonds 
are regarded as building blocks of the chain.\cite{kh} Hence, the division of  the spin-1 chain at one of the bonds corresponds to the division of  the chain between $\v{S}_{2l}$ and $\v{S}_{2l+1}$  along line A in Fig. \ref{fig:cut} (division A) in the present model.  Therefore, we  identify the phase with even ES degeneracy as the SPT phase and  that with odd ES degeneracy as the trivial phase with division A.  Thus, to maintain the correspondence with the spin-1 chain, we have to adopt division A.

Nevertheless, one can also define the ES by the division of the chain between $\v{S}_{2l-1}$ and $\v{S}_{2l}$ along line B in Fig. \ref{fig:cut} (division B). In this case, no virtual free  spins appear between the left and right subsystems in the Haldane phase. In this context, the Haldane phase in our model can also be regarded as a trivial spin-gap phase.

We calculate the ES along the horizontal lines (a)-(d) in Fig. \ref{fig:phase} for both divisions A and B to obtain deeper insight into the ground state of our model. In most cases, the number of the states kept in each subsystem is 240. In the neighborhood of the phase boundaries, it is increased up to 480 to ensure the convergence. Several large eigenvalues $w_{\alpha}$ are shown in Fig. \ref{fig:enta} for division A and in Fig. \ref{fig:entb} for division B. The degeneracies of the largest eigenvalues are also shown in Fig. \ref{fig:enta} as $g_{\rm max}$ for division A and in Fig. \ref{fig:entb} as $\tilde{g}_{\rm max}$ for division B. In all I$_{\sed}$ phases, the relations  $g_{\rm max}=2\sed+1$ and $\tilde{g}_{\rm max}=2\sed$ hold. Physically, this implies that 
the virtual free  spins with magnitude $\sed$  appear for division A, while those with magnitude $\sed-1/2$ appear for division B at the boundary between the left and right subsystems. This confirms that the $I_{\sed}$ phases with integer $\sed$ and half-odd integer $\sed$ are topologically distinct. As for the valence bond structure, this also means that the number of  valence bonds is $2\sed$ across division A, while it is $2\sed-1$  across division B. This is consistent with the valence bond solid picture proposed in Ref. \citen{hts}.

Unfortunately, the convergence of the iDMRG 
 is not good near the phase boundary. Hence, it is difficult to determine the phase boundary using only the iDMRG results.  Although some of the data for the  phase boundaries  taken from Ref. \citen{hts} are slightly inconsistent with the present iDMRG data, these would be attributed to the ambiguity in the size-extrapolation procedures in Ref \citen{hts}.  

\section{Summary and Discussion}
We employed the iDMRG method to calculate  the ES of the spin-1/2 ferromagnetic-antiferromagnetic alternating Heisenberg chain with  next-nearest-neighbour interaction for divisions A and B.   We have confirmed the presence of successive frustration-induced phase transitions between a series of topological and trivial spin-gap phases predicted in Ref. \citen{hts} using the characterization of these phases in terms of  the ES. The valence bond solid picture of each phase is confirmed.

In the frustrated quantum spin chains, various kinds of exotic phases induced by the interplay of quantum fluctuation, frustration, and ferromagnetic correlation are predicted near the frustration-induced transition point between the ferromagnetic and nonmagnetic phases. For example, the spin-nematic phase\cite{Chubukov1991,htsdec,khdlt} and partial ferrimagnetic phases with incommensurate spin correlation\cite{ym,khferri,shimo1,htsf,htsdec,khdlt,shimo2,furuya-giamarchi} have been extensively discussed. The presence of successive transitions between SPT and trivial spin-gap phases 
 is also one of the exotic phenomena in frustrated quantum spin chains.  Recently,  these successive transitions have also been confirmed by the mapping onto the nonlinear $\sigma$ model.\cite{furuya} The possibility of similar successive transitions has been suggested in other frustrated spin chains,\cite{htsdec,khdlt} although the results were not conclusive owing to the limited system size. Further theoretical and experimental searches for these exotic phases would be promising subjects in the field of frustrated quantum magnetism.

The author thanks S. C. Furuya for enlightening comments and discussion on the earlier version of this paper. Part of the numerical computation in this work has been carried out using the facilities of the Supercomputer Center, Institute for Solid State Physics, University of Tokyo, and   Yukawa Institute Computer Facility in Kyoto University. This work is supported by a Grant-in-Aid for Scientific Research (C) (25400389) from the Japan Society for the Promotion of Science.


\begin{thebibliography}{10}
\bibitem{hal1}F. D. M. Haldane,  Phys. Rev. Lett. {\bf 50} 1153 (1983). 
\bibitem{hal2}F. D. M. Haldane,  Phys. Lett. A {\bf 93} 464 (1983). 
\bibitem{Pollmann2010}
F.~Pollmann, A.~M. Turner, E.~Berg, and M.~Oshikawa,  Phys. Rev. B {\bfseries
  81} 064439 (2010).
\bibitem{Pollmann2012}
F.~Pollmann, E.~Berg, A.~M. Turner, and M.~Oshikawa,  Phys. Rev. B {\bfseries
  85} 075125 (2012).
\bibitem{Hirano2008}
T.~Hirano, H.~Katsura, and Y.~Hatsugai,  Phys. Rev. B {\bfseries 77}
  094431 (2008).
\bibitem{Zang2010}
J.~Zang, H.-C. Jiang, Z.-Y. Weng, and S.-C. Zhang,  Phys. Rev. B {\bfseries 81}
 224430  (2010).
\bibitem{Chen2011} X. Chen, Z.-C. Gu, and X.-G. Wen,  
Phys. Rev. B {\bfseries 83} 035107 (2011).
\bibitem{hts}
K. Hida, K. Takano, and H. Suzuki,  J. Phys. Soc. Jpn. {\bf 82} 064703 (2013). 
\bibitem{dmitriev} D. V. Dmitriev, V. Ya. Krivnov, and A. A. Ovchinnikov,  Phys. Rev. B {\bf 56} 5985 (1997).
\bibitem{dmitriev2} D. V. Dmitriev, V. Ya. Krivnov, and A. A. Ovchinnikov ,  Eur. Phys. J. B {\bf 14} 91 (2000). 

\bibitem{McCulloch2008}
I. P. McCulloch, arXiv:0804.2509
\bibitem{Schllwock2011} U. Schollw\"ock,   Ann. Phys. {\bf 326} 96 (2011).
\bibitem{kh} K. Hida,  Phys. Rev. B {\bf 45} 2207 (1992).
\bibitem{khdlt} K. Hida,  J. Phys. Soc. Jpn. {\bf 77} 044707 (2008), [Errata {\bf 79} 028001 (2010)]. 
\bibitem{htsdec} K. Hida and K. Takano,  Phys. Rev. B {\bf 78} 064407 (2008).
\bibitem{Chubukov1991}
A.~V. Chubukov,  Phys. Rev. B {\bfseries 44} 4693 (1991).
\bibitem{htsf} K. Hida, K. Takano, and H. Suzuki,  J. Phys. Soc. Jpn. {\bf 79} 114703 (2010).
\bibitem{ym} S. Yoshikawa and S. Miyashita,  J. Phys. Soc. Jpn. Suppl. {\bf 74} 71 (2005).
\bibitem{khferri} K. Hida,  J. Phys.: Condens. Matter {\bf 19} 145225  (2007).
\bibitem{shimo1} T. Shimokawa and H. Nakano,  J. Phys. Soc. Jpn. {\bf 80} 043703 (2011); 
J. Phys. Soc. Jpn. {\bf 80} 125003 (2011); 
 J. Phys. Soc. Jpn. {\bf 81} 084710 (2012).
\bibitem{shimo2} T. Shimokawa and H. Nakano,  J. Korean Phys. Soc. {\bf 63}, 591 (2013).
\bibitem{furuya-giamarchi} S. C. Furuya and T. Giamarchi,  Phys. Rev. B {\bf 89}, 205131 (2014).
\bibitem{furuya} S. C. Furuya,  private communication.

\end{thebibliography}
\end{document}